\documentclass[journal]{IEEEtran}
\usepackage{amsmath,graphicx}
\usepackage{amsmath,amssymb,epsfig,psfrag,cite,subfigure}
\include{macros}
\usepackage{graphicx}
\usepackage{balance}
\usepackage{epstopdf}
\usepackage[font=footnotesize]{caption}
\usepackage{bm}

\usepackage{geometry}
\usepackage{tcolorbox}
\usepackage{float}

\usepackage{soul}

\usepackage{amsthm}

\usepackage{nicefrac}

\usepackage{lipsum,multicol}

\usepackage{algorithm}
\usepackage{algpseudocode}

\usepackage{enumitem}
\usepackage{mwe}    
\usepackage{epsfig,psfrag}
\usepackage{subfigure}
\usepackage{color}
\usepackage{url}
\usepackage{mathtools,xparse}

\usepackage{gensymb}

\usepackage[multiple]{footmisc}

\usepackage{xcolor}

\usepackage{pgfplots}
\usepackage{amsmath}
\usepackage{amsfonts}
\usepackage{bm}
\usetikzlibrary{arrows.meta}
\tikzset{every picture/.style={line width=0.6pt}}
\usetikzlibrary{calc,positioning}
\usetikzlibrary{arrows.meta,
                backgrounds,
                chains,
                fit,
                quotes}

\theoremstyle{remark}

\newtheoremstyle{mytheoremstyle} 
    {\topsep}                    
    {\topsep}                    
    {\upshape}                   
    {.5em}                           
    {\itshape}                   
    {.}                          
    {.5em}                       
    {}  
    
\theoremstyle{mytheoremstyle}

\newtheoremstyle{iremark}
  {\topsep}   
  {\topsep}   
  {\upshape}  
  {0.2in}       
  {\itshape}  
  {.}         
  {5pt plus 1pt minus 1pt} 
  {\thmname{#1}\thmnumber{ \itshape#2}\thmnote{ (#3)}} 

\theoremstyle{iremark}

\newtheorem*{proof}{Proof}

\DeclareFontFamily{U}{mathx}{\hyphenchar\font45}
\DeclareFontShape{U}{mathx}{m}{n}{
	<5> <6> <7> <8> <9> <10>
	<10.95> <12> <14.4> <17.28> <20.74> <24.88>
	mathx10
}{}
\DeclareSymbolFont{mathx}{U}{mathx}{m}{n}
\DeclareFontSubstitution{U}{mathx}{m}{n}

\DeclarePairedDelimiter\abs{\lvert}{\rvert}%

\makeatletter \renewcommand\d[1]{\ensuremath{%
		\;\mathrm{d}#1\@ifnextchar\d{\!}{}}}
\makeatother

\makeatletter
\newcommand*\rel@kern[1]{\kern#1\dimexpr\macc@kerna}
\newcommand*\widebar[1]{%
  \begingroup
  \def\mathaccent##1##2{%
    \rel@kern{0.8}%
    \overline{\rel@kern{-0.8}\macc@nucleus\rel@kern{0.2}}%
    \rel@kern{-0.2}%
  }%
  \macc@depth\@ne
  \let\math@bgroup\@empty \let\math@egroup\macc@set@skewchar
  \mathsurround\z@ \frozen@everymath{\mathgroup\macc@group\relax}%
  \macc@set@skewchar\relax
  \let\mathaccentV\macc@nested@a
  \macc@nested@a\relax111{#1}%
  \endgroup
}

\newcommand{\yy}{\mathbf{y}}

\newcommand{\nn}{\mathbf{n}}

\newcommand{\ssb}{\mathbf{s}}

\newcommand{\fb}{\mathbf{f}}

\newcommand{\FF}{\mathbf{F}}

\newcommand{\snr}{{\rm{SNR}}}

\newcommand{\hermit}{\mathsf{H}}

\usepackage{accents}

\newcommand{\atantwo}{{\rm{atan2}}}

\newcommand{\rmtx}{{\rm{Tx}}}

\newcommand{\Ntx}{N_\rmtx}

\newcommand{\atx}{\mathbf{a}}

\newcommand{\complexset}[2]{ \mathbb{C}^{#1 \times #2}  }

\newcommand{\mtCN}{{\mathcal{CN}}}

\renewcommand{\baselinestretch}{0.96}

\graphicspath{{./Figures/}}

\usepackage[nolist]{acronym}
\begin{acronym}[ACRONYM]
\acro{AE}{autoencoder}
\acro{AI}{artificial intelligence}
\acro{AoA}{angle-of-arrival}
\acro{AoD}{angle-of-departure}
\acro{BS}{base station}
\acro{BP}{belief propagation}
\acro{CDF}{cumulative density function}
\acro{CRB}{Cram\'er-Rao bound}
\acro{DA}{data association}
\acro{D-MIMO}{distributed multiple-input multiple-output}
\acro{DL}{downlink}
\acro{EM}{electromagnetic}
\acro{E2E}{end-to-end}
\acro{FIM}{Fisher information matrix}
\acro{GDOP}{geometric dilution of precision}
\acro{GNSS}{global navigation satellite system}
\acro{GPS}{global positioning system}
\acro{HWI}{hardware impairment}
\acro{IP}{incidence point}
\acro{IQ}{in-phase and quadrature}
\acro{ISAC}{integrated sensing and communication}
\acro{ICI}{inter-carrier interference}
\acro{JCS}{Joint Communication and Sensing}
\acro{JRC}{joint radar and communication}
\acro{JRC2LS}{joint radar communication, computation, localization, and sensing}
\acro{IMU}{inertial measurement unit}
\acro{IOO}{indoor open office}
\acro{IoT}{Internet of Things}
\acro{IRN}{infrastructure reference node}
\acro{KPI}{key performance indicator}
\acro{LoS}{line-of-sight}
\acro{LS}{least-squares}
\acro{MC}{mutual coupling}
\acro{MCRB}{misspecified Cram\'er-Rao bound}
\acro{MIMO}{multiple-input multiple-output}
\acro{MISO}{multiple-input single-output}
\acro{ML}{maximum likelihood}
\acro{mmWave}{millimeter-wave}
\acro{MSE}{mean squared error}
\acro{NLoS}{non-line-of-sight}
\acro{NN}{neural network}
\acro{NR}{new radio}
\acro{OFDM}{orthogonal frequency-division multiplexing}
\acro{OTFS}{orthogonal time-frequency-space}
\acro{OEB}{orientation error bound}
\acro{PEB}{position error bound}
\acro{VEB}{velocity error bound}
\acro{PRS}{positioning reference signal}
\acro{QoS}{Quality of Service}
\acro{RAN}{radio access network}
\acro{RAT}{radio access technology}
\acro{RCS}{radar cross section}
\acro{RedCap}{reduced capacity}
\acro{RF}{radio frequency}
\acro{RIS}{reconfigurable intelligent surface}
\acro{RFS}{random finite set}
\acro{RMSE}{root mean squared error}
\acro{RTK}{real-time kinematic}
\acro{RTT}{round-trip-time}
\acro{SLAM}{simultaneous localization and mapping}
\acro{SLAT}{simultaneous localization and tracking}
\acro{SNR}{signal-to-noise ratio}
\acro{ToA}{time-of-arrival}
\acro{TDoA}{time-difference-of-arrival}
\acro{TR}{time-reversal}
\acro{TXRX}[TX/RX]{transmitter/receiver}
\acro{Tx}{transmitter}
\acro{Rx}{receiver}
\acro{UE}{user equipment}
\acro{UL}{uplink}
\acro{ULA}{uniform linear array}
\acro{UWB}{ultra wideband}
\acro{XL-MIMO}{extra-large MIMO}
\acro{NLL}{negative log-likelihood}
\end{acronym}
\begin{document}

\bstctlcite{IEEEexample:BSTcontrol}

\title{Spatial Signal Design for Positioning via End-to-End Learning}
\author{Steven~Rivetti,
        José~Miguel~Mateos-Ramos,~\IEEEmembership{Student Member,~IEEE},
        Yibo~Wu,~\IEEEmembership{Student Member,~IEEE},
        Jinxiang~Song,~\IEEEmembership{Student Member,~IEEE}, 
        Musa~Furkan~Keskin,~\IEEEmembership{Member,~IEEE},
        Vijaya~Yajnanarayana,~\IEEEmembership{Senior Member,~IEEE}, 
        Christian~Häger,~\IEEEmembership{Member,~IEEE},
        Henk~Wymeersch,~\IEEEmembership{Senior Member,~IEEE}
       \thanks{
This work was supported by the European Commission through the H2020 project Hexa-X (Grant Agreement no.~101015956), the Swedish Research Council (grant no. 2020-04718), MSCA-IF grant
888913 (OTFS-RADCOM) and by a grant from the Chalmers AI Research Centre Consortium.}
        \thanks{The authors (except V.~Yajnanarayana) are with the Department
of Electrical Engineering, Chalmers University of Technology, 41258 Gothenburg, Sweden (e-mail: henkw@chalmers.se). Y.~Wu is also with Ericsson Research, Gothenburg, Sweden. V.~Yajnanarayana is with Ericsson Research, India. } \vspace{-5mm}}

\maketitle

\begin{abstract}
    This letter considers the problem of \ac{E2E} learning for joint optimization of transmitter precoding and receiver processing for mmWave downlink positioning. Considering a \ac{MISO} scenario, we propose a novel \ac{AE} architecture to estimate \ac{UE} position with multiple \acp{BS} and demonstrate that \ac{E2E} learning can match model-based design, both for \ac{AoD} and position estimation, under ideal conditions without model deficits and outperform it in the presence of hardware impairments.
    
    
\end{abstract}

\vspace{-5mm}
\acresetall
\section{Introduction}\label{sec_intro}

\IEEEPARstart{T}{he} combination of high delay resolution at mmWave frequencies thanks to large bandwidth and high angular resolution thanks to large arrays is an important  enabler for accurate positioning in 5G \cite{TR38.855} and beyond \cite{b5g_commag_2021}. The estimation of \ac{ToA}, \ac{AoA},  and \acf{AoD} is enabled by \emph{designed pilot signals} in time, frequency, and in space (at the \acf{BS}) \cite{keating2019overview}. Such designs, in combination with advanced signal processing, can leverage the physical resources efficiently when suitable models are available. 
Traditionally, signal designs were optimized for broadcast performance in order to localize all users irrespective of their position \cite{dwivedi2021positioning}. 
Recently, there has been an increased focus on spatial per-user signal design, leveraging a priori knowledge of the user's location in order to further improve  accuracy, both for positioning \cite{precoderNil2018} and sensing \cite{liu2018toward}. Signal designs can be categorized as \emph{model-based} \cite{fascista2020low,precoderNil2018,successiveLocBF_2019,tasos_precoding2020,signalDesign_TVT_2022} or  \emph{based on \ac{AI}} \cite{e2e_radar_TAES_2022,learnProbingmmWave_2022,Liu2022isac,mateos2021end}. 
Model-based signal designs can be performed based on simple heuristics \cite{fascista2020low}, or  on minimizing the \ac{CRB} on the \ac{AoA}, \ac{AoD}, or the position via the \ac{PEB}. After relaxation, the optimization problems can be cast in convex forms, leading to elegant and efficient designs (e.g.,  \cite{precoderNil2018} for angle estimation and \cite{signalDesign_TVT_2022} for positioning). From these solutions, online adaptive precoders \cite{successiveLocBF_2019} and 
robust designs based on predetermined codebooks with power allocation \cite{tasos_precoding2020} have been considered. 

An important limitation of model-based designs is that they require a model of the transmitter, receiver, and propagation channel. Under model mismatch, e.g., \acp{HWI}, model-based approaches may exhibit degraded performance. Moreover, in certain cases, even with perfect model knowledge, finding optimal signal designs can be intractable. To remedy these two shortcomings,
\ac{E2E} learning has been  gaining interest, first in the context of communication  \cite{Oshea2017intro} and more recently for sensing \cite{e2e_radar_TAES_2022}, but not for positioning. The principle is to model the entire system as an \ac{AE} \cite{Oshea2017intro} or by a combination or a reinforcement learning transmitter and a supervised learning receiver \cite{Aoudia2018alternating},
combined with a suitable loss function (see, e.g., \cite{e2e_radar_TAES_2022}).
An application of \ac{E2E} learning for spatial precoder design can be found in  \cite{learnProbingmmWave_2022}, where the probing codebook is implemented by a \ac{NN} module that is jointly trained with the beam predictor 
in order to predict the optimal narrow beam.
Furthermore, \cite{Liu2022isac} extends the learned beamforming to \ac{ISAC} by implementing the transmitter as a convolutional \ac{NN} able to learn the features of historical channel and predict the next beamforming matrix. \Ac{E2E} learning in the presence of \acp{HWI} for \ac{ISAC} has been proposed in \cite{mateos2021end}. 
\ac{AI}-based solutions have also been applied in other forms to deal with \acp{HWI}, e.g., \cite{chen2022sdoanet} proposes a super-resolution direction of arrival network, implemented as a convolutional \ac{NN}, that can outperform \ac{AoA} estimation methods under \ac{MC}.

\begin{figure}
\centering
\includegraphics[width=0.9\linewidth]{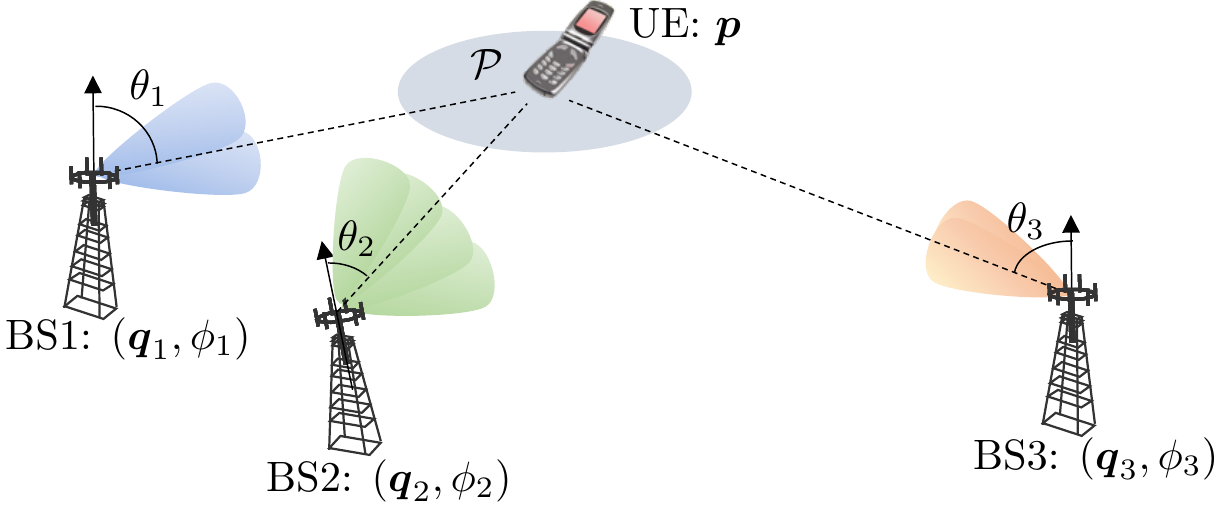}
\caption{The MISO downlink scenario comprising $I>1$ multiple-antenna \acp{BS} and a single-antenna \ac{UE}. The \ac{UE}  determines its position based on the \ac{AoD} estimates with respect to the \acp{BS}.} \label{fig:setup} 
\vspace{-5mm}
\end{figure}


In this paper, \ac{E2E} learning is applied for the first time in positioning, in order to jointly optimize transmit beamformers and receiver-side algorithms, even in the presence of \acp{HWI}. Our contributions are \textit{(i)} a novel \ac{AE} architecture and loss function for \ac{AoD}- and positioning-optimized signal design and estimator design;
\textit{(ii)} a detailed performance comparison to a state-of-the-art model-based benchmark and corresponding \acp{CRB}; \textit{(iii)} an evaluation under different \acp{HWI}, namely array element inter-distance perturbation and array \ac{MC}, demonstrating the robustness of the proposed \ac{E2E} solution.

\section{System Model}\label{sec_sysmodel}

\subsection{Scenario and Signal Model}
We consider a mmWave \ac{MISO} downlink  scenario with $I>1$ multiple-antenna \acp{BS} and a single-antenna \ac{UE} with unknown location $\bm{p} = [p_1 \ p_2]^T \in \mathcal{P} \subset \mathbb{R}^2$, where $\mathcal{P}$ is the prior location information.
Each \ac{BS} $i$ has a known location $\bm{q}_i = [q_{i,1} \ q_{i,2}]^T \in \mathbb{R}^{2}$ and orientation $\psi_i \in [-\frac{\pi}{2},\frac{\pi}{2}]$, and is assumed to be equipped with an $\Ntx$-element \ac{ULA} with $\lambda/2$ antenna spacing, where $\lambda$ denotes the wavelength of the carrier. The scenario is visualized in Fig.~\ref{fig:setup}.

The $i$-th BS broadcasts a narrowband signal over $T>1$ successive transmissions. We assume that \ac{BS} transmissions are orthogonalized in time or frequency \cite{dwivedi2021positioning}, leading to the observation at the \ac{UE} from \ac{BS} $i$ at transmission $t$ given by
\begin{align}\label{eq_ym}
      y_{i,t} = \alpha_i \, \atx^\top(\theta_i) \fb_{i,t} s_{i,t} + n_{i,t}~,
\end{align}
where $s_{i,t}$ denotes the pilot signal with an unit power $ \abs{s_{i,t}}^2  = 1$,
 $\alpha_i \in \mathbb{C}$ and $\theta_i \in [-\frac{\pi}{2},+\frac{\pi}{2}]$ denote, respectively, the complex channel gain and \ac{AoD} from the $i$-th BS, $\atx(\theta_i) \in \mathbb{C}^{\Ntx}$ is the array steering vector at the BS (ULA of $\Ntx$ elements and $\lambda/2$ antenna spacing), $\fb_{i,t} \in \mathbb{C}^{\Ntx}$ is the precoder employed by the $i$-th BS at time $t$, and $n_{i,t} \sim \mtCN(0, \sigma^2)$ is the additive white noise with variance $\sigma^2$, accounting also for the signal energy. In a more compact form, \eqref{eq_ym} can be rewritten as
\begin{align}\label{eq_yi}
   \yy_i =  \alpha_i \, (\FF_i^{\top} \atx(\theta_i)) \odot \ssb_i + \nn_i ~,
\end{align}
where $\odot$ represents the Hadamard product, 
$\yy_i = [y_{i,1} \ldots y_{i,T}]^{\top}$, $ \FF_i = \left[ \fb_{i,1} \, \ldots \, \fb_{i,T} \right] \in \complexset{\Ntx}{T}$ is the precoder matrix of the $i$-th BS, $\ssb_i = [s_{i,1} \ldots s_{i,T}]^{\top}$, and $\nn_i = [n_{i,1} \ldots n_{i,T}]^{\top}$.
From the \ac{UE} and \ac{BS} positions, the \ac{AoD} is computed as 
\begin{align} 
    \label{atan}
     \theta_i &= \atantwo( p_2 - q_{i,2} , p_1 - q_{i,1} )-\psi_i~,
\end{align}
which accounts for the \ac{BS} orientation. We assume that the UE lies in the angular sector $\mathcal{U}_i=[\theta_{i,\min}, \theta_{i,\max}]\in \mathbb{R}^2$ with respect to the \ac{BS} $i$, depending on the uncertainty region $\mathcal{P}$.

\begin{figure}
\centering
\begin{tikzpicture}[font=\scriptsize, >=stealth,nd/.style={draw,circle,inner sep=0pt,minimum size=0pt}, 
blk/.style={draw,minimum height=1cm,text width=1.5cm, text centered},  
blkrcv/.style={draw,minimum height=1cm,text width=1.7cm, text centered}, 
blkthr/.style={draw,minimum height=0.8cm,text width=1cm, text centered}, 
x=0.6cm,y=0.55cm]

\tikzset{threshold_fig/.pic={
\draw (-1,0)--(1,0);
\draw[red,line width=1.0 pt] (-1,-1)--(0,-1)--(0,1)--(1,1);
}}

\path		
	(-10,0)coordinate[](input1){} 
    node(beamformer1)[blk, right=1 of input1, fill=green!20]{Beamformer \\
	$f_{\epsilon_i}$} 

    node(Channel1)[blk, right=1.5 of beamformer1]{Wireless channel}
    
    node(aod)[blkrcv, right=1.5 of Channel1, fill=blue!20]{AoD estimator\\
$f_{\mu}$} 
   
    coordinate[right=1 of aod](aod_out1){}

;

\draw[->] (input1)--node[above]{$\boldsymbol{\xi}_i$}(beamformer1);
\draw[->] (beamformer1)--node[above]{$\FF_i$}(Channel1);
\draw[->] (Channel1)--node[above]{$\yy_i$}(aod);
\draw[->] (aod)--node[above]{$\hat{\theta}_i$}(aod_out1);
\end{tikzpicture}
\caption{Block diagram of the proposed \ac{AE} architecture dedicated to \ac{AoD} estimation, the green and blue blocks are implemented as trainable feed-forward NNs.}\vspace{-3mm}
\label{NN aod}
\end{figure}
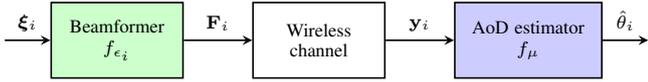


\subsection{Hardware Impairment Models}
Without \acp{HWI}, the steering vectors are given by 
$[\atx(\theta)]_k=e^{j \pi k \sin(\theta)}$, $k=0,\ldots, \Ntx-1$. We now describe the impact of inter-antenna element spacing perturbations and \ac{MC}, which lead to an impaired steering vector, denoted by $\tilde{\atx}(\theta)$. 
\subsubsection{Antenna Element Spacing Perturbations} \label{sec:AES}
We introduce the vector of inter-element distances as $\bm{d} \in \mathbb{R}^{\Ntx-1}$, where without HWIs, $\bm{d} =\frac{\lambda}{2} \bm{1}_{\Ntx-1}$. Here, $\bm{1}_{\Ntx-1}$ denotes a vector of $(\Ntx-1)$ ones. With spacing perturbations caused by HWIs \cite{yassine2022fl}, the distance is modeled by
\begin{align}\label{distance noise}
    \bm{d} =\frac{\lambda}{2} \bm{1}_{\Ntx-1}+ \bm{\gamma}, \gamma_{k} \sim \mathcal{N}(0,\sigma_\lambda^2),
\end{align}
so that the perturbed steering vector becomes $[\tilde{\atx}(\theta)]_{k}=e^{j 2 \pi k (d_{k}/\lambda)\sin(\theta)}$.

\begin{figure}
\centering
\begin{tikzpicture}[font=\scriptsize, >=stealth,nd/.style={draw,circle,inner sep=0pt,minimum size=0pt}, 
blk/.style={draw,minimum height=1cm,text width=1cm, text centered},  
blkrcv/.style={draw,minimum height=1cm,text width=1.2cm, text centered}, 
blkthr/.style={draw,minimum height=0.8cm,text width=1cm, text centered}, 
x=0.6cm,y=0.55cm]

\tikzset{threshold_fig/.pic={
\draw (-1,0)--(1,0);
\draw[red,line width=1.0 pt] (-1,-1)--(0,-1)--(0,1)--(1,1);
}}

\path		
	(-10,0)coordinate[](input1){} 
node(beamformer1)[blkrcv, right=1 of input1, fill=green!20]{Beamformer \\
 	$f_{\epsilon_1}$} 

node(Channel1)[blk, right=1 of beamformer1]{Wireless channel}

coordinate[below=3.5 of input1](inputI){} 
node(beamformerI)[blkrcv, right=1 of inputI, fill=green!20]{Beamformer \\
 	$f_{\epsilon_I}$} 

node(ChannelI)[blk, right=1 of beamformerI]{Wireless channel}
    

coordinate[right=1 of Channel1](corner1){} 
coordinate[right=1 of ChannelI](cornerI){} 
coordinate[below=2 of corner1](concat){} 
	
    
	
	
node(dot)[below =0.05 of beamformer1,font=\Large]{\vdots}
node(dot2)[below=0.05 of Channel1,font=\Large]{\vdots}
	


node(pos)[blkrcv, right=2.5 of concat, fill=red!20]{POS estimator\\
 	$f_{\beta}$} 
 coordinate[right=1.5 of pos](pos_out){}


    
   coordinate[right=1 of pos](pos_out){}
    

    
    
    
   
    

;

\draw[->] (input1)--node[above]{$\boldsymbol{\xi}_1$}(beamformer1);
\draw[->] (beamformer1)--node[above]{$\FF_1$}(Channel1);

\draw[->] (inputI)--node[above]{$\boldsymbol{\xi}_I$}(beamformerI);
\draw[->] (beamformerI)--node[above]{$\FF_I$}(ChannelI);

\draw[-] (Channel1)--node[above]{$\yy_1$}(corner1);
\draw[-] (ChannelI)--node[above]{$\yy_I$}(cornerI);

\draw[-] (corner1)--node[above]{}(concat);
\draw[-] (cornerI)--node[above]{}(concat);
\draw[->] (concat)--node[above]{$[\yy_1^\top\ldots\yy_I^\top]$}(pos);

\draw[->] (pos)--node[above]{$\hat{\bm{p}}$}(pos_out);
\end{tikzpicture}
\caption{Block diagram of the proposed \ac{AE} architecture dedicated to position estimation, the red block is implemented as trainable feed-forward NN.}\vspace{-5mm}
\label{NN pos}
\end{figure}
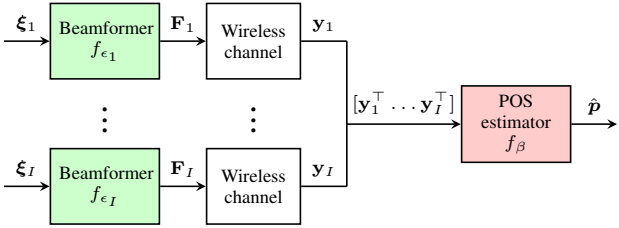
\subsubsection{Mutual Coupling}\label{sec:MC}

Following \cite{Zheng2017coupling}, we introduce a coupling matrix $\bm{B} \in\complexset{\Ntx}{\Ntx}$, which is modeled as a banded symmetric Toeplitz matrix whose entries are collected in the vector $\bm{c}=[1,c_1,\ldots,c_M]^\top~(0<\abs{c_M}<\ldots<\abs{c_1}<1)$, where $M$ is the number of half-wavelength increments for which the MC contribution is assumed non-negligible,
so that $\tilde{\atx}(\theta)=\bm{B}{\atx}(\theta)$.




\section{End-to-End Learning}
In this section, we describe the proposed architectures, the associated loss functions, and the model-based benchmark. 
\subsection{End-to-End Learning Architecture}

We consider two separate \ac{AE} architectures for \ac{AoD} and position estimation, as shown in Fig.~\ref{NN aod} and Fig.~\ref{NN pos}, respectively. Fig.~\ref{NN aod} shows an \ac{E2E} architecture to learn \ac{BS} precoder design (highlighted in green) and \ac{UE}-side \ac{AoD} estimation from each \ac{BS} (highlighted in blue). Fig.~\ref{NN pos} shows an \ac{E2E} architecture to learn \ac{BS} precoder design (highlighted in green) and \ac{UE}-side position estimation, based on the combined observation from all \acp{BS} (highlighted in red). We assume the wireless channel blocks are instantaneously differentiable.

\subsubsection{Precoder \ac{NN}}
Each BS has its own precoder. The precoder for BS $i$ is implemented by an \ac{NN} $f_{\epsilon_i}:\mathbb{R}^3 \rightarrow \mathbb{C}^{\Ntx T}$, with learnable parameters $\epsilon_i$. Instead of directly using the AoD uncertainty region $\mathcal{U}_i$ as the NN input, we find it helpful to feed an over-determined parameterization of $\mathcal{U}_i$ as $\boldsymbol{\xi}_i \in \mathbb{R}^3$, with
\begin{align}
    \boldsymbol{\xi}_i=[\theta_{i,\min},\theta_{i,\max},(\theta_{i,\max}-\theta_{i,\min})/2]^\top.
\end{align}
The \ac{NN} output is a real-valued vector with a size $\mathbb{R}^{2\Ntx T}$ that is then converted into the complex-valued precoding matrix $\FF_i \in \complexset{\Ntx}{T}$. In this conversion, complex numbers are obtained by concatenating the real and imaginary parts, followed by a normalization with its Frobenius norm. 
\subsubsection{\ac{AoD} Estimation \ac{NN}}
The \ac{AoD} estimator for \ac{BS} $i$ is implemented by another \ac{NN} $f_\mu:\mathbb{C}^T \rightarrow \mathbb{R}$, with learnable parameters $\mu$, which takes the observation $\yy_i$ as the input  and generates an estimate $\hat{\theta}_i$. Since the \ac{AoD} estimation  process is identical for each \ac{BS}, 
all $I$ \ac{AoD} estimators share the same parameters.

\subsubsection{Position Estimation \ac{NN}}


The position estimator is implemented as $f_{\beta}:\mathbb{C}^{I\times T} \rightarrow \mathbb{R}^{2}$, with  learnable parameters $\beta$, which takes as input $\yy = [\yy_1^{\top},\yy_2^{\top} ~\ldots, \yy_I^{\top}]^{\top}$ and generates as output the position estimate $\hat{\bm{p}}$.  
Since AoD and position are intrinsically related, this direct  approach could potentially be replaced with a two-step solution, by leveraging the AoD estimation NNs, at a cost of possible performance loss (due to the data processing theorem), but with possibly lower complexity. A two-step solution also necessitates computing the AoD uncertainties, as in \cite{mateos2021end}.

\subsection{Loss Functions}
The \ac{E2E} \ac{AoD} estimation and \ac{E2E} position estimation require two dedicated loss functions:
\begin{itemize}
    \item \emph{\ac{AoD} estimation loss:} The loss function is the \ac{MSE} between the estimated and true AoDs:
    \begin{align}\label{rmse aod}
 \mathcal{L}_{\text{AoD}}(\epsilon_i,\mu)= \mathbb{E}\{ \vert \hat{\theta}_i-\theta_i \vert^2\}~. 
\end{align} 
The \ac{AoD} estimators at the \ac{UE} corresponding to each BS share \ac{NN} parameters, so there is no need for separate training. 
Since the \acp{AoD} are limited to $[-\pi/2,\pi/2)$, there is no risk of wrapping effects, making the \ac{MSE} meaningful in this scenario. 
    \item \emph{Positioning loss:} The loss function is set to 
\begin{align}\label{rmse pos}
 \mathcal{L}_{\text{position}}(\epsilon_1,\ldots,\epsilon_I,\beta)= \mathbb{E}\{\Vert  \hat{\boldsymbol{p}} -\boldsymbol{p} \Vert^2_2 \}~.      
\end{align}
\end{itemize}


\subsection{Benchmarks}
As a comparison, each of the \acp{NN} in Fig.~\ref{NN aod} and Fig.~\ref{NN pos} will be evaluated against a state-of-the-art benchmark. 

\subsubsection{Transmit Precoding Benchmark}
The chosen precoder matrix for the BS $i$ is a heuristic solution to the problem of minimization of worst-case \ac{CRB} on \ac{AoD} estimation over the uncertainty region $\mathcal{U}_i$. It consists of a hybrid base codebook, comprising both directional beams and their derivatives \cite{signalDesign_TVT_2022} 
\begin{align}\label{F_digital}
\FF_i^{\text{heur}}&=[\FF_i^{\text{dir}},~ \FF_i^{\text{der}}] \in\complexset{\Ntx}{T} ~,\\
\FF_i^{\text{dir}}&=[\atx(\theta_{i,0}), \dots,\atx(\theta_{i,T/2})]\in \complexset{\Ntx}{(T/2)}~,\\
\FF_i^{\text{der}}&=[\dot{\atx}(\theta_{i,0}),\dots,\dot{\atx}(\theta_{i,T/2})]\in \complexset{\Ntx}{(T/2)}~,
\end{align}
where $\{\theta_{i,g}\}_{g=1}^{T/2}$ represents the evenly spaced angular grid in $\mathcal{U}_i$ and $\dot{\atx}(\theta)=\partial \atx(\theta)/ \partial \theta $. The benchmark precoder $\FF_i^\text{b}$, defined as $=[\sqrt{\rho_1}\fb_{i,1}^\text{heur}, \ldots, \sqrt{\rho_{T}}\fb_{i,T}^\text{heur}]$ where $\fb_{i,t}^\text{heur}$ denotes the $t$-th column of $\FF_i^{\text{heur}}$, is obtained by finding the power allocation vector $\bm{\rho}=[\rho_1 \ldots \rho_T]^\top$ that minimizes the \ac{CRB} on \ac{AoD} estimation \cite{signalDesign_TVT_2022}. Then, $\FF_i^\text{b}$ is normalized to have unit Frobenius norm; the same operation is implemented by the normalization layer at the output of the beamformer NN, ensuring the usage of the same total power between the two approaches. 


\subsubsection{\ac{AoD} Estimation Benchmark}
The \ac{UE} implements  \ac{ML} estimation, based on  \eqref{eq_yi}, yielding \cite{Fascista2019downlinkPos} 
\begin{align}\label{ml}
    \hat{\theta}_i^\text{b}=\arg\min_{\theta_i \in \mathcal{U}_i} \frac{\lvert \yy_i^\hermit \FF_i^{\top} \atx(\theta_i) \rvert^2 }{ \Vert \FF_i^{\top} \atx(\theta_i) \Vert^2} ~.
\end{align}

\subsubsection{Position Estimation Benchmark}
Given the \ac{AoD} estimates from \eqref{ml}, we formulate the measurement likelihood $p(\hat{\theta}_i^{\text{b}}|\theta_i)$ as $p(\hat{\theta}_i^{\text{b}}|\theta_i)\propto \exp(-(\hat{\theta}_i^{\text{b}}-\theta_i)^2/(2 \sigma_i^2) )$, where 
%
 $\sigma^2_i$ can be obtained from the \ac{CRB} of the \ac{AoD} estimator at \ac{BS} $i$. Then, it immediately follows that the \ac{ML} estimator is
\begin{align}\label{ml pos}
    \hat{\boldsymbol{p}}^{\text{b}}=\text{arg} \min_{\boldsymbol{p} \in \mathcal{P}}~ \sum_{i=1}^I 
    \frac{1}{2 \sigma_i^2} (\hat{\theta}_i^{\text{b}}+\psi_i-\atantwo(\boldsymbol{q_i},\boldsymbol{p}))^2~.
\end{align}

\section{Simulation Results}

\subsection{Simulation Parameters}\label{sec:parameters}
We consider a scenario with $I=2$ \acp{BS}, located at $\bm{q}_1=[-5,0]$ and $\bm{q}_2=[3,0]$ with orientations $\bm{\psi}=[0\degree,10\degree]$,  each with $\Ntx =32$ antenna elements. The number of transmissions is set to 
$T = 20$ with pilots $s_{i,t}=1$, 
and the width of $\mathcal{U}_i$ varies uniformly between $10^\circ$ and $20^\circ$. 
The channel gains are set based on a target \ac{SNR}, i.e., $\snr_i = \abs{\alpha_i}^2/\sigma^2$, and the \acp{SNR} range from $-5$ dB to $30$ dB. The phase of $\alpha_i$ is uniformly distributed in $[0,2\pi]$ and the wavelength is set to $10.7$ mm  (corresponding to a carrier frequency of $28$ GHz).  



For modeling the \acp{HWI}, we generate the 
\ac{MC} matrix  $\bm{B}$ as a banded symmetric Toeplitz matrix built from the coefficients vector 
$\boldsymbol{c}=[1,0.9e^{-j\pi/3},0.75e^{j\pi/4},0.55e^{-j\pi/10},0.25e^{-j\pi/6}]^\top$, while for generating the antenna element spacing perturbations, we set $\sigma_\lambda=\lambda/100$. 

\begin{table} 
\caption{ \ac{NN} structures.}
\begin{center}
\begin{tabular}{c c c c} 
 \hline
 Network & Input layer & Hidden layers & Output layer \\ 
 \hline
 Beamformer $f_\epsilon$ & 3 & H,H,H,H,H,H & $ \Ntx T$ (linear)\\ 
 \ac{AoD} decoder $f_\mu$  & 2T & H,H,H,H,2H,2H & 1 (tanh)\\ 
 POS decoder $f_\beta$  & 4T & H,H,H,H,2H,2H & 2 (linear)\\ 
 \hline
\end{tabular}
\end{center}\label{tab:NNparameters}
\vspace{-5mm}
\end{table}

\subsection{Autoencoder Training}
The mini-batch size $S$ is set to $10000$ and we train with mean \ac{AoD} uniformly distributed in $[-60\degree,60\degree]$ and  $\mathcal{U}_i$'s width uniformly distributed within $[10\degree,20\degree]$. 
In terms of positioning \ac{AE}, the training follows a similar rationale: each minibatch's sample is associated to a true position $\boldsymbol{p}$, modelled as a $2$-D uniform random variable  within a $10~\text{m}^2$ area in front of the \acp{BS}. 
The observations $\yy_i$ are then generated by calculating $\theta_i$ according to  \eqref{atan}. Then, the mean of $\mathcal{U}_i$ is set to $\theta_{\text{mid},i}=\theta_i+\nu_i$, where $\nu$ is a random variable varying uniformly within the interval $[-15\degree,15\degree]$, as the a priori information  induces a $30\degree$ wide $\mathcal{U}_i$ on both \ac{BS}.
The beam former NN input is then defined as $\bm{\xi}_i=[\theta_{\text{mid},i}-15\degree,\theta_{\text{mid},i}+15\degree,15\degree]^\top$. 

Based on a hyper-parameter search, which aimed to determine the smallest NN with the best possible performance, the number of hidden neurons `H' is set to $256$ and each layer uses a rectified linear unit (ReLU) activation function.
Further details are provided in Table \ref{tab:NNparameters}. 
We also note that in practical applications, it may be of interest to use NN architectures with less complexity (i.e., fewer layers and/or neurons per layer) by sacrificing some accuracy.
In terms of optimizer, we use the Adam optimizer \cite{kingma2014adam} with a learning rate controlled by a scheduler whose starting value is $0.001$ and lower bound is at $10^{-8}$.
We have found that re-training the systems with fixed \ac{SNR} ranging from $-5$ dB to $30$ dB yields better results than using a different \ac{SNR} in every batch or sample.

\subsection{Results}
\subsubsection{Without Hardware Impairments}
\begin{figure}
    \centering
    \subfigure[{Precoder responses with \ac{SNR}=$10$ dB}]{\input{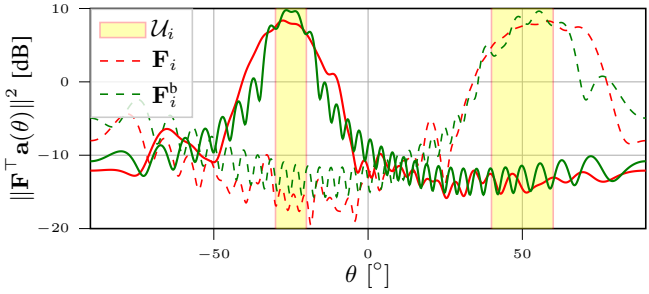}
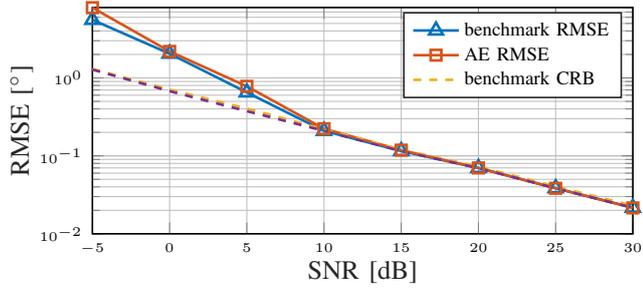
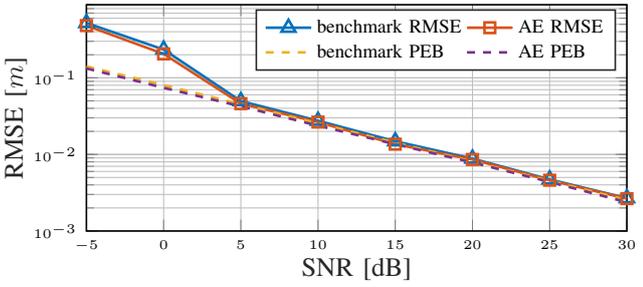}
    \subfigure[{AoD RMSE for $\mathcal{U}_1=[40\degree,60\degree]$}]{
%
%
\definecolor{mycolor1}{rgb}{0.00000,0.44700,0.74100}%
\definecolor{mycolor2}{rgb}{0.85000,0.32500,0.09800}%
\definecolor{mycolor3}{rgb}{0.92900,0.69400,0.12500}%
\definecolor{mycolor4}{rgb}{0.49400,0.18400,0.55600}%
\begin{tikzpicture}

\begin{axis}[%
width=0.8\linewidth,
height=3cm,
at={(0.758in,0.485in)},
scale only axis,
legend columns=1,
xmin=-5,
xmax=30,
xlabel style={font=\color{white!15!black}},
xlabel={SNR [dB]},
ymode=log,
ymin=0.01,
ymax=8,
yminorticks=true,
ylabel style={yshift=-0.3cm},
xlabel style={yshift=0.3cm},
ylabel style={font=\color{white!15!black}},
ylabel={RMSE [$^{\circ}$]},
axis background/.style={fill=white},
xmajorgrids,
ymajorgrids,
yminorgrids,
legend style={legend cell align=left, align=left, draw=white!15!black,font=\small,nodes={scale=0.8, transform shape}},
tick label style={font=\tiny}
]
\addplot [color=mycolor1, line width=1.0pt, mark size=3.0pt, mark=triangle, mark options={solid,mycolor1}]
  table[row sep=crcr]{%
-5	5.5659323410898\\
0	2.047260325967\\
5	0.6520900044444\\
10	0.209349775072591\\
15	0.115219682950237\\
20	0.0698513223280066\\
25	0.03843413570909426\\
30	0.0216349775072591\\
};
\addlegendentry{benchmark RMSE}

\addplot [color=mycolor2, line width=1.0pt, mark size=2.0pt, mark=square, mark options={solid, mycolor2}]
  table[row sep=crcr]{%
-5	7.966947632891991\\
0	2.189581960866722\\
5	0.780458943820841\\
10	0.222813947232446\\
15	0.118322431324213\\
20	0.069932432456781\\
25	0.038375346576786\\
30	0.021478126986471\\
};
\addlegendentry{AE RMSE}

\addplot [color=mycolor3, line width=1.0pt,dashed]
  table[row sep=crcr]{%
-5	1.30219682950237\\
0	0.7088513223280066\\
5	0.4043413570909426\\
10	0.219349775072591\\
15	0.120219682950237\\
20	0.0728513223280066\\
25	0.04043413570909426\\
30	0.0229349775072591\\
};
\addlegendentry{benchmark CRB}

\addplot [color=mycolor4, line width=1.0pt,dashed]
  table[row sep=crcr]{%
-5	1.27854398549\\
0	0.67590449869\\
5	0.3743413570909426\\
10	0.209349775072591\\
15	0.115219682950237\\
20	0.0698513223280066\\
25	0.03843413570909426\\
30	0.0216349775072591\\
};

\end{axis}
\end{tikzpicture}%
}
    \subfigure[{Positioning RMSE for $\bm{p}=[0.5,5]~\text{m}$}]{
%
%
\definecolor{mycolor1}{rgb}{0.00000,0.44700,0.74100}%
\definecolor{mycolor2}{rgb}{0.85000,0.32500,0.09800}%
\definecolor{mycolor3}{rgb}{0.92900,0.69400,0.12500}%
\definecolor{mycolor4}{rgb}{0.49400,0.18400,0.55600}%
\begin{tikzpicture}

\begin{axis}[%
width=0.8\linewidth,
height=3cm,
at={(0.758in,0.485in)},
scale only axis,
legend columns=2,
xmin=-5,
xmax=30,
xlabel style={font=\color{white!15!black}},
xlabel={SNR [dB]},
ymode=log,
ymin=0.001,
ymax=0.9,
yminorticks=true,
ylabel style={yshift=-0.3cm},
xlabel style={yshift=0.3cm},
ylabel style={font=\color{white!15!black}},
ylabel={RMSE [${m}$]},
axis background/.style={fill=white},
xmajorgrids,
ymajorgrids,
yminorgrids,
legend style={legend cell align=left, align=left, draw=white!15!black,font=\small,nodes={scale=0.8, transform shape}},
tick label style={font=\tiny}
]
\addplot [color=mycolor1, line width=1.0pt, mark size=3.0pt, mark=triangle, mark options={solid, mycolor1}]
  table[row sep=crcr]{%
-5	0.520029708681001\\
0	0.234004010918135\\
5	0.0496446613170299\\
10	0.0275028219423505\\
15	0.0148504050649719\\
20	0.008810715602050892\\
25	0.004712004432399142\\
30	0.00268209533818779\\
};
\addlegendentry{benchmark RMSE}

\addplot [color=mycolor2, line width=1.0pt, mark size=2.0pt, mark=square, mark options={solid, mycolor2}]
  table[row sep=crcr]{%
-5	0.480938974136\\
0	0.20533625632\\
5	0.0460118594088168\\
10	0.0264356612403256\\
15	0.0136076280900236\\
20	0.00858852872060672\\
25	0.00460195034463936\\
30	0.002646372948736\\
};
\addlegendentry{AE RMSE}

\addplot [color=mycolor3, line width=1.0pt,dashed]
  table[row sep=crcr]{%
-5	0.14090045858947\\
0	0.0802967436123108\\
5	0.0451541772114166\\
10	0.0253920598509479\\
15	0.0142790045858947\\
20	0.00802967436123108\\
25	0.00451541772114166\\
30	0.00253920598509479\\
};
\addlegendentry{benchmark PEB}

\addplot [color=mycolor4, line width=1.0pt,dashed]
  table[row sep=crcr]{%
-5	0.13273366447096\\
0	0.074096342899847\\
5	0.042410802704892\\
10	0.02372257052016\\
15	0.0135873366447096\\
20	0.0078483428998472\\
25	0.0043730802704896\\
30	0.00237272257052\\
};
\addlegendentry{AE PEB}

\end{axis}
\end{tikzpicture}%
e}
    \caption{Results without \acp{HWI}: Performance comparison of the \acp{AE} with the benchmark.}
    \label{no imp}\vspace{-3mm}
\end{figure}

Fig.~\ref{no imp}-(a) shows the aggregated response $\Vert \FF^{\top} \atx(\theta) \Vert^2$ of the \ac{AoD}-optimized learned precoder $\FF = \FF_i$ for the two BSs for angle uncertainty intervals $\mathcal{U}_1=[40\degree,60\degree]$ and $\mathcal{U}_2=[-30\degree,-20\degree]$, along with that of the benchmark precoder $\FF = \FF_i^\text{b}$. 
Despite the \ac{AE} having no knowledge of the benchmark precoder, the learned precoder has a strong similarity in terms of the aggregate response\footnote{The position-optimized precoders exhibit similar trends (results not shown for space reasons).}. 
%
Fig.~\ref{no imp}-(b) shows the \ac{AoD} \ac{RMSE} performance vs.~\ac{SNR} for BS $1$, along with the corresponding \acp{CRB}\footnote{The benchmark CRB and the AE CRB are computed by employing $\FF_i^\text{b}$ and $\FF_i$ as the precoding matrices, respectively \cite[Ch.~3]{kay1993fundamentals}. Since the CRB depends on the transmit signal, not on the receiver processing, different precoders may lead to different CRB values. Additionally, we note that under HWIs the steering vector model used in the CRB computation is the true one, i.e., $\tilde{\atx}(\theta)$.}. The implicit power allocation process carried out by the \ac{AE} in finding $\FF_i$ is able to achieve the same performance bounds obtained through the explicit optimization process to determine $\FF_i^\text{b}$.
Furthermore, both approaches are able to attain the \ac{CRB} at an \ac{SNR} around $10$ dB.
This trend is confirmed for positioning as well: Fig.~\ref{no imp}-(c) shows that the \ac{E2E} solution can reach the same \ac{PEB} as its model-based counterpart, attaining it around an \ac{SNR} of $5$ dB.

\subsubsection{Results under Hardware Impairments}
Next, we show the impact of model mismatch caused by \acp{HWI} on the \ac{AoD} and position estimation, while revealing the capability of the proposed \ac{AE} to compensate for the resulting performance degradation. 

\begin{figure}
    \centering
%
%
\definecolor{mycolor1}{rgb}{0.00000,0.44700,0.74100}%
\definecolor{mycolor2}{rgb}{0.85000,0.32500,0.09800}%
\definecolor{mycolor3}{rgb}{0.92900,0.69400,0.12500}%
\definecolor{mycolor4}{rgb}{0.49400,0.18400,0.55600}%
\begin{tikzpicture}[every node/.append style={font=\small}]

\begin{axis}[%
width=0.8\linewidth,
height=3cm,
at={(0.758in,0.485in)},
scale only axis,
xmin=1,
xmax=5,
xlabel style={font=\color{white!15!black}},
xtick={1,2,3,4,5},
xticklabels={${\lambda}/{30}$,${\lambda}/{50}$,${\lambda}/{100}$,${\lambda}/{200}$,${\lambda}/{300}$},
xlabel={$\sigma_\lambda$},
ymode=log,
ymin=0.003,
ymax=0.5,
yminorticks=true,
x dir=reverse,
ylabel style={font=\color{white!15!black}},
ylabel style={yshift=-0.3cm},
ylabel={RMSE [${m}$]},
axis background/.style={fill=white},
xmajorgrids,
ymajorgrids,
yminorgrids,
legend columns=2,
legend style={legend cell align=left, align=left, draw=white!15!black,font=\small,nodes={scale=0.8, transform shape}},
tick label style={font=\tiny}
]
\addplot [color=mycolor1, dashed, line width=1.0pt, mark size=3.0pt, mark=triangle, mark options={solid, mycolor1}]
  table[row sep=crcr]{%
1	0.3033\\
2	0.2013\\
3	0.0979\\
4	0.0481\\
5	0.028\\
};
\addlegendentry{benchmark RMSE}

\addplot [color=mycolor2, line width=1.0pt, mark size=2.0pt, mark=square, mark options={solid, mycolor2}]
  table[row sep=crcr]{%
1	0.007901454\\
2	0.007734334\\
3	0.007489378\\
4	0.007431274\\
5	0.007484931\\
};
\addlegendentry{AE RMSE}

\addplot [color=mycolor3, line width=1.0pt,dashed]
  table[row sep=crcr]{%
1	0.0077\\
2	0.0077\\
3	0.0077\\
4	0.0077\\
5	0.0077\\
};
\addlegendentry{benchmark PEB}
\addplot [color=mycolor4, line
width=1.0pt,dashed]
  table[row sep=crcr]{%
1	0.0074\\
2	0.0074\\
3	0.0074\\
4	0.0074\\
5	0.0074\\
};
\addlegendentry{AE PEB}

\end{axis}
\end{tikzpicture}%
    \caption{Positioning RMSE performances at an SNR of $20$ dB for increasingly large array element spacing perturbations $\sigma_\lambda$.}
    \vspace{-3mm}
    \label{peb vs sigma}
\end{figure}
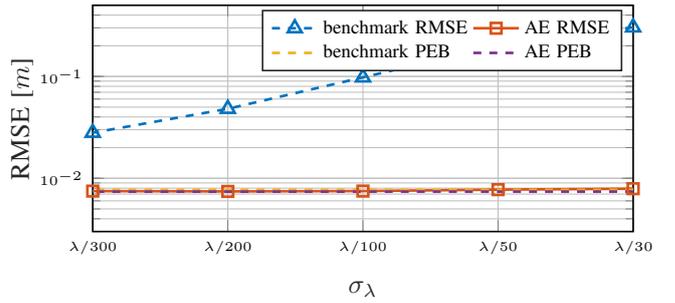

First, we consider array element spacing perturbations, shown in Fig.~\ref{precoder fluctuations} and Fig.~\ref{rmse fluctuations}. 
The observations are generated using the model from Section \ref{sec:AES}, while the model-based benchmark is unaware of this impairment. From Fig.~\ref{precoder fluctuations}, we observe that the \ac{AE} precoder responses are less similar to the benchmark, compared to the case without \ac{HWI}: this difference in the precoders can be interpreted as an active adaptation to $\tilde{\atx}(\theta)$. This is also seen in Fig.~\ref{rmse fluctuations}, which shows the \ac{AoD} and positioning \ac{RMSE}. In particular, at medium and high \ac{SNR} values, the benchmark suffers from significant performance penalties due to mismatch between the true model $\tilde{\atx}(\theta)$ and the employed model $\atx(\theta)$, in line with the theoretical results from \cite{Chen-HWI-2022}. The \ac{AE} is able to attain its \ac{CRB} in both \ac{AoD} and position estimation, verifying the effectiveness of the proposed architecture under model imperfections.
Moreover, Fig.~\ref{peb vs sigma} plots the position RMSE with respect to $\sigma_\lambda$ for a fixed \ac{SNR} of $20$ dB, which further confirms the robustness of the \ac{E2E} solution. Specifically, the positioning \ac{AE} can achieve the \ac{PEB} regardless of $\sigma_\lambda$, whereas the model-based approach leads to a performance penalty that increases with $\sigma_\lambda$. 

\begin{figure}
    \centering
    \input{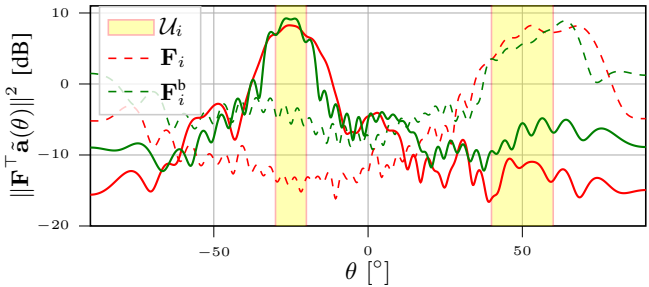}
    \caption{Aggregate response of the precoders $\FF_i$ compared against the respective benchmarks $\FF_i^\text{b}$ under array element spacing perturbations with $\sigma_\lambda=\lambda/100$ and for $\text{SNR}=10$ dB.}
    \label{precoder fluctuations}\vspace{-3mm}
\end{figure}
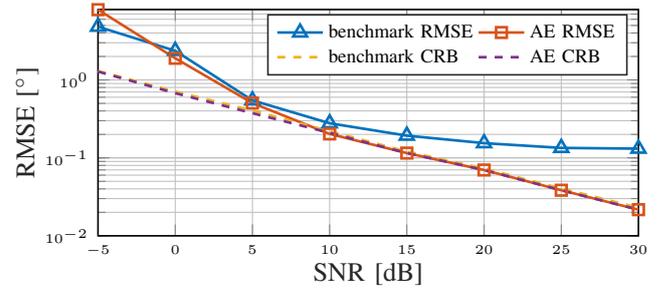
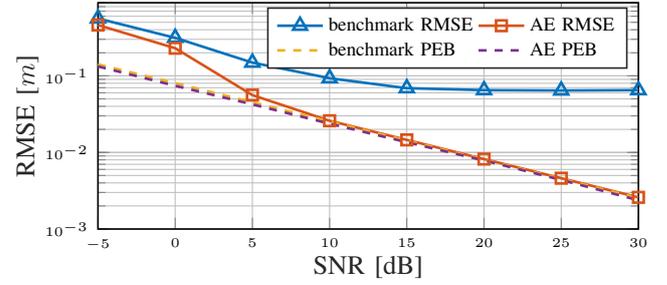
\begin{figure}
    \centering
    \subfigure[{AoD RMSE for $\mathcal{U}_1=[40\degree,60\degree]$}]{
%
%
\definecolor{mycolor1}{rgb}{0.00000,0.44700,0.74100}%
\definecolor{mycolor2}{rgb}{0.85000,0.32500,0.09800}%
\definecolor{mycolor3}{rgb}{0.92900,0.69400,0.12500}%
\definecolor{mycolor4}{rgb}{0.49400,0.18400,0.55600}%
\begin{tikzpicture}[every node/.append style={font=\small}]

\begin{axis}[%
width=0.8\linewidth,
height=3cm,
at={(0.758in,0.485in)},
scale only axis,
xmin=-5,
xmax=30,
xlabel style={font=\color{white!15!black}},
xlabel={SNR [dB]},
ymode=log,
ymin=0.01,
ymax=8,
yminorticks=true,
ylabel style={font=\color{white!15!black}},
ylabel={RMSE [$^{\circ}$]},
axis background/.style={fill=white},
ylabel style={yshift=-0.3cm},
xlabel style={yshift=0.3cm},
xmajorgrids,
ymajorgrids,
yminorgrids,
legend columns=2,
legend style={legend cell align=left, align=left, draw=white!15!black,font=\small,nodes={scale=0.8, transform shape}},
tick label style={font=\tiny}
]
\addplot [color=mycolor1, line width=1.0pt, mark size=3.0pt, mark=triangle, mark options={solid, mycolor1}]
  table[row sep=crcr]{%
-5	4.81851213403301\\
0	2.36372022939895\\
5	0.546878995209047\\
10	0.277918810226166\\
15	0.192790443872886\\
20	0.154436269201992\\
25	0.13418417414233\\
30	0.131263429510336\\
};
\addlegendentry{benchmark RMSE}

\addplot [color=mycolor2, line width=1.0pt, mark size=2.0pt, mark=square, mark options={solid, mycolor2}]
  table[row sep=crcr]{%
-5	7.96694763289199\\
0	1.8958196086672\\
5	0.5045894382084\\
10	0.202813947232446\\
15	0.115219131312344\\
20	0.069854352346142\\
25	0.038435256266532\\
30	0.0216349775072591\\
};
\addlegendentry{AE RMSE}

\addplot [color=mycolor3, line width=1.0pt,dashed]
  table[row sep=crcr]{%
-5	1.30219682950237\\
0	0.7088513223280066\\
5	0.4043413570909426\\
10	0.219349775072591\\
15	0.120219682950237\\
20	0.0728513223280066\\
25	0.04043413570909426\\
30	0.0229349775072591\\
};
\addlegendentry{benchmark CRB}

\addplot [color=mycolor4, line width=1.0pt,dashed]
  table[row sep=crcr]{%
-5	1.27854398549\\
0	0.67590449869\\
5	0.3743413570909426\\
10	0.209349775072591\\
15	0.115219682950237\\
20	0.0698513223280066\\
25	0.03843413570909426\\
30	0.0216349775072591\\
};
\addlegendentry{AE CRB}

\end{axis}
\end{tikzpicture}%
}
    \subfigure[{Positioning RMSE for $\bm{p}=[0.5,5]~\text{m}$}]{
%
%
\definecolor{mycolor1}{rgb}{0.00000,0.44700,0.74100}%
\definecolor{mycolor2}{rgb}{0.85000,0.32500,0.09800}%
\definecolor{mycolor3}{rgb}{0.92900,0.69400,0.12500}%
\definecolor{mycolor4}{rgb}{0.49400,0.18400,0.55600}%
\begin{tikzpicture}[every node/.append style={font=\small}]

\begin{axis}[%
width=1\linewidth,
height=6.5cm,
width=0.8\linewidth,
height=3cm,
at={(0.758in,0.485in)},
scale only axis,
xmin=-5,
xmax=30,
xlabel style={font=\color{white!15!black}},
xlabel={SNR [dB]},
ymode=log,
ymin=0.001,
ymax=0.9,
yminorticks=true,
ylabel style={font=\color{white!15!black}},
ylabel={RMSE [${m}$]},
axis background/.style={fill=white},
xmajorgrids,
ymajorgrids,
yminorgrids,
ylabel style={yshift=-0.3cm},
xlabel style={yshift=0.3cm},
legend columns=2,
legend style={legend cell align=left, align=left, draw=white!15!black,font=\small,nodes={scale=0.8, transform shape}},
tick label style={font=\tiny}
]
\addplot [color=mycolor1, line width=1.0pt, mark size=3.0pt, mark=triangle, mark options={solid, mycolor1}]
  table[row sep=crcr]{%
-5	0.560706160275510\\
0	0.313295570978282\\
5	0.148829772093182\\
10	0.093006851759749\\
15	0.068959322266018\\
20	0.065025974887054\\
25	0.064387099164054\\
30	0.065000833359958\\
};
\addlegendentry{benchmark RMSE}

\addplot [color=mycolor2, line width=1.0pt, mark size=2.0pt, mark=square, mark options={solid, mycolor2}]
  table[row sep=crcr]{%
-5	0.460608321602369\\
0	0.22941966922428\\
5	0.056069930145100\\
10	0.025907025569222\\
15	0.014568591090337\\
20	0.008192520819896\\
25	0.004606993014510\\
30	0.002590702556922\\
};
\addlegendentry{AE RMSE}

\addplot [color=mycolor3, line width=1.0pt,dashed]
  table[row sep=crcr]{%
-5	0.14090045858947\\
0	0.0802967436123108\\
5	0.0451541772114166\\
10	0.0253920598509479\\
15	0.0142790045858947\\
20	0.00802967436123108\\
25	0.00451541772114166\\
30	0.00253920598509479\\
};
\addlegendentry{benchmark PEB}

\addplot [color=mycolor4, line width=1.0pt,dashed]
  table[row sep=crcr]{%
-5	0.13273366447096\\
0	0.074096342899847\\
5	0.042410802704892\\
10	0.02372257052016\\
15	0.0135873366447096\\
20	0.0078483428998472\\
25	0.0043730802704896\\
30	0.00237272257052\\
};
\addlegendentry{AE PEB}
\end{axis}
\end{tikzpicture}%
}
    \caption{RMSE performance assessment in the presence of array element spacing perturbations.}
    \label{rmse fluctuations}\vspace{-3mm}
\end{figure} 

\begin{figure}
    \centering
    \input{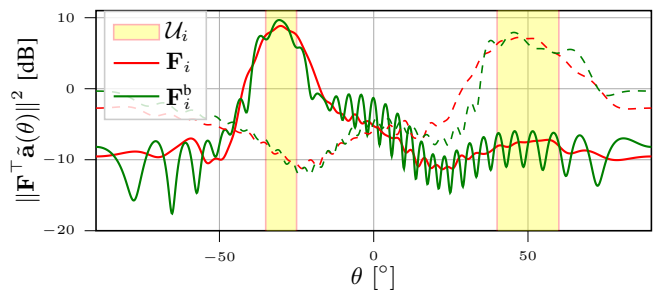}
    \caption{
    Aggregate response of the precoders $\FF_i$ compared against the respective benchmarks $\FF_i^\text{b}$ under \ac{MC} for $\text{SNR}=10$ dB.}
    \label{precoder mc}\vspace{-3mm}
\end{figure}

\vspace{10mm}

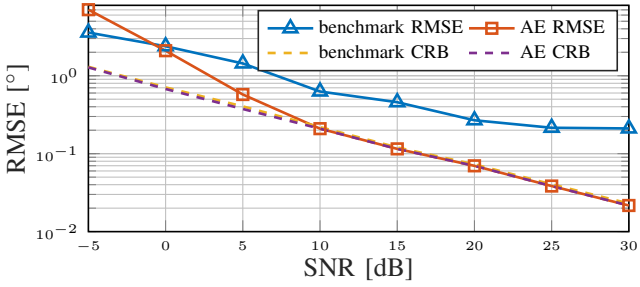
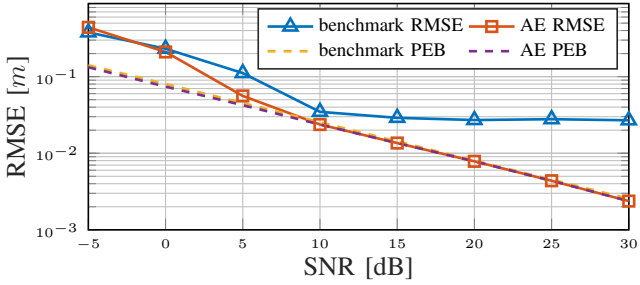
\begin{figure}
    \centering
    \subfigure[{AoD RMSE for $\mathcal{U}_1=[40\degree,60\degree]$}]{
%
%
\definecolor{mycolor1}{rgb}{0.00000,0.44700,0.74100}%
\definecolor{mycolor2}{rgb}{0.85000,0.32500,0.09800}%
\definecolor{mycolor3}{rgb}{0.92900,0.69400,0.12500}%
\definecolor{mycolor4}{rgb}{0.49400,0.18400,0.55600}%

\begin{tikzpicture}

\begin{axis}[%
width=0.8\linewidth,
height=3cm,
at={(0.758in,0.485in)},
scale only axis,
xmin=-5,
xmax=30,
xlabel style={font=\color{white!15!black}},
xlabel={SNR [dB]},
ymode=log,
ymin=0.01,
ymax=8,
yminorticks=true,
ylabel style={font=\color{white!15!black}},
ylabel={RMSE [$^{\circ}$]},
axis background/.style={fill=white},
ylabel style={yshift=-0.3cm},
xlabel style={yshift=0.3cm},
xmajorgrids,
ymajorgrids,
yminorgrids,
legend columns=2,
legend style={legend cell align=left, align=left, draw=white!15!black,font=\small,nodes={scale=0.8, transform shape}},
tick label style={font=\tiny}
]
\addplot [color=mycolor1, line width=1.0pt, mark size=3.0pt, mark=triangle, mark options={solid, mycolor1}]
  table[row sep=crcr]{%
-5	3.58044707111289\\
0	2.3875758735593\\
5	1.43753393234724\\
10	0.627552334270315\\
15	0.457326403427969\\
20	0.268261345116908\\
25	0.214940831116165\\
30	0.210711434233994\\
};
\addlegendentry{benchmark RMSE}

\addplot [color=mycolor2, line width=1.0pt, mark size=2.0pt, mark=square, mark options={solid, mycolor2}]
  table[row sep=crcr]{%
-5	7.037934464048582\\
0	2.094700930059104\\
5	0.5743413570909426\\
10	0.209349775072591\\
15	0.115219682950237\\
20	0.0698513223280066\\
25	0.03843413570909426\\
30	0.0216349775072591\\
};
\addlegendentry{AE RMSE}

\addplot [color=mycolor3, line width=1.0pt,dashed]
  table[row sep=crcr]{%
-5	1.30219682950237\\
0	0.7088513223280066\\
5	0.4043413570909426\\
10	0.219349775072591\\
15	0.120219682950237\\
20	0.0728513223280066\\
25	0.04043413570909426\\
30	0.0229349775072591\\
};
\addlegendentry{benchmark CRB}

\addplot [color=mycolor4, line width=1.0pt,dashed]
  table[row sep=crcr]{%
-5	1.27854398549\\
0	0.67590449869\\
5	0.3743413570909426\\
10	0.209349775072591\\
15	0.115219682950237\\
20	0.0698513223280066\\
25	0.03843413570909426\\
30	0.0216349775072591\\
};
\addlegendentry{AE CRB}
\end{axis}
\end{tikzpicture}%
}\\
    \subfigure[{Positioning RMSE for $\bm{p}=[0.5,5]~\text{m}$}]{
%
%
\definecolor{mycolor1}{rgb}{0.00000,0.44700,0.74100}%
\definecolor{mycolor2}{rgb}{0.85000,0.32500,0.09800}%
\definecolor{mycolor3}{rgb}{0.92900,0.69400,0.12500}%
\definecolor{mycolor4}{rgb}{0.49400,0.18400,0.55600}%
\begin{tikzpicture}

\begin{axis}[%
width=0.8\linewidth,
height=3cm,
at={(0.758in,0.485in)},
scale only axis,
legend columns=2,
xmin=-5,
xmax=30,
xlabel style={font=\color{white!15!black}},
xlabel={SNR [dB]},
ymode=log,
ymin=0.001,
ymax=0.9,
yminorticks=true,
ylabel style={yshift=-0.3cm},
xlabel style={yshift=0.3cm},
ylabel style={font=\color{white!15!black}},
ylabel={RMSE [${m}$]},
axis background/.style={fill=white},
xmajorgrids,
ymajorgrids,
yminorgrids,
legend style={legend cell align=left, align=left, draw=white!15!black,font=\small,nodes={scale=0.8, transform shape}},
tick label style={font=\tiny}
]
\addplot [color=mycolor1, line width=1.0pt, mark size=3.0pt, mark=triangle, mark options={solid, mycolor1}]
  table[row sep=crcr]{%
-5	0.378290954782097\\
0	0.231513819417864\\
5	0.111247134344121\\
10	0.034706089407461\\
15	0.028981309069218\\
20	0.027133555472634\\
25	0.027835448690330\\
30	0.026924610373371\\
};
\addlegendentry{benchmark RMSE}

\addplot [color=mycolor2, line width=1.0pt, mark size=2.0pt, mark=square, mark options={solid, mycolor2}]
  table[row sep=crcr]{%
-5	0.440608321602369\\
0	0.2101966922428\\
5	0.05601103769827147\\
10	0.02372257052016\\
15	0.0135873366447096\\
20	0.0078483428998472\\
25	0.0043730802704896\\
30	0.00237272257052\\
};
\addlegendentry{AE RMSE}

\addplot [color=mycolor3, line width=1.0pt,dashed]
  table[row sep=crcr]{%
-5	0.14090045858947\\
0	0.0802967436123108\\
5	0.0451541772114166\\
10	0.0253920598509479\\
15	0.0142790045858947\\
20	0.00802967436123108\\
25	0.00451541772114166\\
30	0.00253920598509479\\
};
\addlegendentry{benchmark PEB}

\addplot [color=mycolor4, line width=1.0pt,dashed]
  table[row sep=crcr]{%
-5	0.13273366447096\\
0	0.074096342899847\\
5	0.042410802704892\\
10	0.02372257052016\\
15	0.0135873366447096\\
20	0.0078483428998472\\
25	0.0043730802704896\\
30	0.00237272257052\\
};
\addlegendentry{AE PEB}

\end{axis}
\end{tikzpicture}%
}
    \caption{RMSE performance assessment in presence of MC.}
    \label{rmse mc}
\end{figure}
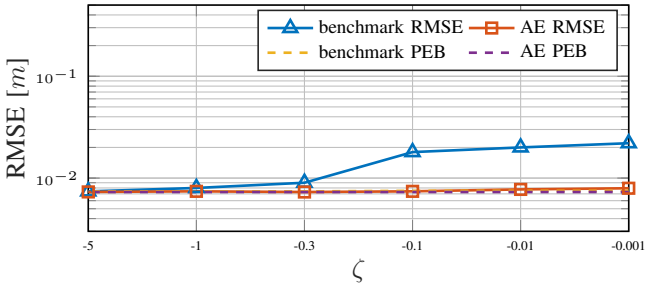
\begin{figure}
    \centering
%
%
\definecolor{mycolor1}{rgb}{0.00000,0.44700,0.74100}%
\definecolor{mycolor2}{rgb}{0.85000,0.32500,0.09800}%
\definecolor{mycolor3}{rgb}{0.92900,0.69400,0.12500}%
\definecolor{mycolor4}{rgb}{0.49400,0.18400,0.55600}%
\begin{tikzpicture}[every node/.append style={font=\small}]

\begin{axis}[%
width=0.8\linewidth,
height=3cm,
at={(0.758in,0.485in)},
scale only axis,
xmin=1,
xmax=6,
xlabel style={font=\color{white!15!black}},
xtick={1,2,3,4,5,6},
xticklabels={-0.001, -0.01, -0.1, -0.3, -1, -5},
xlabel={$\zeta$},
x dir=reverse,
ymode=log,
ymin=0.003,
ymax=0.5,
yminorticks=true,
ylabel style={font=\color{white!15!black}},
ylabel style={yshift=-0.3cm},
xlabel style={yshift=0.3cm},
ylabel={RMSE [${m}$]},
axis background/.style={fill=white},
xmajorgrids,
ymajorgrids,
yminorgrids,
legend columns=2,
legend style={legend cell align=left, align=left, draw=white!15!black,font=\small,nodes={scale=0.8, transform shape}},
tick label style={font=\tiny}
]
\addplot [color=mycolor1, line width=1.0pt, mark size=3.0pt, mark=triangle, mark options={solid, mycolor1}]
  table[row sep=crcr]{%
1	0.022\\
2	0.020\\
3	0.018\\
4	0.009\\
5	0.008\\
6	0.0074\\
};
\addlegendentry{benchmark RMSE}

\addplot [color=mycolor2, line width=1.0pt, mark size=2.0pt, mark=square, mark options={solid, mycolor2}]
  table[row sep=crcr]{%
1	0.0079483319\\
2	0.00774322131\\
3	0.007412431323\\
4	0.0073001243\\
5	0.007413425423\\
6	0.007313425423\\
};
\addlegendentry{AE RMSE}

\addplot [color=mycolor3, line width=1.0pt,dashed]
  table[row sep=crcr]{%
1	0.0075\\
2	0.0075\\
3	0.0075\\
4	0.0074\\
5	0.0074\\
6	0.0074\\
};
\addlegendentry{benchmark PEB}

\addplot [color=mycolor4, line width=1.0pt,dashed]
  table[row sep=crcr]{%
1	0.0073\\
2	0.0073\\
3	0.0073\\
4	0.0073\\
5	0.0073\\
6	0.0073\\
};
\addlegendentry{AE PEB}

\end{axis}
\end{tikzpicture}%
    \caption{Positioning RMSE performances at an SNR of $20$ dB for different magnitudes of MC.}
    \label{peb vs mc}
\end{figure}

\vspace{-10mm}
Second, we evaluate the impact of \ac{MC}, where the 
observations are now generated according to the model from  Section \ref{sec:MC}. 
In Fig.~\ref{precoder mc}, the precoder responses are shown, which suggests that the proposed learning-based approach can naturally adapt its precoder to deal with \acp{HWI}, leading to a beampattern that is different from that of the model-based approach.
Fig.~\ref{rmse mc} illustrates the RMSEs and the CRBs of the considered strategies under the impact of \ac{MC}. It is seen that the precoder generated by the \ac{AE} can achieve the same performance bound as the model-based benchmark.
In terms of \ac{RMSE}, the \ac{MC} prevents the benchmark estimator from attaining its bound, while the proposed \ac{AE} can successfully reach the theoretical limits. For positioning with the benchmark estimator, the MC induces an error floor effect beyond $10$ dB, as expected from \cite{Chen-HWI-2022}.
Further insights into the effects of MC are provided in Fig.~\ref{peb vs mc} (using the same RMSE scale as Fig.~\ref{peb vs sigma}), where we model the MC coupling coefficients vector as $|{c}_k|=\exp(\zeta k), k\in \{0,\ldots,4\}$ and retain the phase of the original $\bm{c}$ reported in Section \ref{sec:parameters}. The 
resulting matrix $\bm{B}$ is normalized to have the same Frobenius norm as the matrix $\bm{B}$ built from the vector $\bm{c}$ reported in Section~\ref{sec:parameters}. Similar to Fig.~\ref{peb vs sigma}, the \ac{E2E} solution is able to attain its performance bound, whereas the model-based solution shows a performance penalty inversely proportional to the decay parameter $\zeta$. Comparing with Fig.~\ref{peb vs sigma}, we do however note that the impact of MC is less severe than array spacing perturbations. 





\vspace{-3mm}
\section{Conclusions}
We have addressed the problem of positioning and \ac{AoD} estimation at a \ac{UE}, based on downlink MISO transmission. To this end, we propose a novel \ac{AE} architecture with judiciously designed inputs and loss functions, which jointly learns optimized precoders and receivers under \ac{UE} location uncertainty.
We have compared the \ac{AE} performance against model-based precoder designs and \ac{ML} estimators. 
Through numerical simulations, the learned precoders are shown to yield the same bounds as their model-based counterparts. Without model imperfections, the learned receiver can attain the same RMSE level as the \ac{ML} estimator. In the presence of \acp{HWI}, the learned receiver can significantly outperform the \ac{ML} estimator, especially at high SNRs and large degree of inter-element perturbations and \ac{MC}, showcasing the robustness of the proposed \ac{AE} architecture against model deficits. Possible future work include extension to 3D scenarios and investigation of two-step architectures that exploit the relation between \ac{AoD} and position (i.e., \eqref{ml pos}) to jointly design their corresponding NN estimators for reduced complexity.





\begin{thebibliography}{10}
\providecommand{\url}[1]{#1}
\csname url@samestyle\endcsname
\providecommand{\newblock}{\relax}
\providecommand{\bibinfo}[2]{#2}
\providecommand{\BIBentrySTDinterwordspacing}{\spaceskip=0pt\relax}
\providecommand{\BIBentryALTinterwordstretchfactor}{4}
\providecommand{\BIBentryALTinterwordspacing}{\spaceskip=\fontdimen2\font plus
\BIBentryALTinterwordstretchfactor\fontdimen3\font minus
  \fontdimen4\font\relax}
\providecommand{\BIBforeignlanguage}[2]{{%
\expandafter\ifx\csname l@#1\endcsname\relax
\typeout{** WARNING: IEEEtran.bst: No hyphenation pattern has been}%
\typeout{** loaded for the language `#1'. Using the pattern for}%
\typeout{** the default language instead.}%
\else
\language=\csname l@#1\endcsname
\fi
#2}}
\providecommand{\BIBdecl}{\relax}
\BIBdecl

\bibitem{TR38.855}
3rd Generation Partnership Project~(3GPP), ``Study on {NR} positioning support
  {TR} 38.855,'' \emph{Technical Specification Group Radio Access Network},
  2019.

\bibitem{b5g_commag_2021}
S.~Bartoletti \emph{et~al.}, ``Positioning and sensing for vehicular safety
  applications in {5G} and beyond,'' \emph{IEEE Communications Magazine},
  vol.~59, no.~11, pp. 15--21, 2021.

\bibitem{keating2019overview}
R.~Keating \emph{et~al.}, ``Overview of positioning in {5G} new radio,'' in
  \emph{IEEE International Symposium on Wireless Communication Systems
  (ISWCS)}, 2019, pp. 320--324.

\bibitem{dwivedi2021positioning}
S.~Dwivedi \emph{et~al.}, ``Positioning in {5G} networks,'' \emph{IEEE
  Communications Magazine}, vol.~59, no.~11, pp. 38--44, 2021.

\bibitem{precoderNil2018}
N.~{Garcia} \emph{et~al.}, ``Optimal precoders for tracking the {AoD} and {AoA}
  of a {mmWave} path,'' \emph{IEEE Transactions on Signal Processing}, vol.~66,
  no.~21, pp. 5718--5729, Nov 2018.

\bibitem{liu2018toward}
F.~Liu \emph{et~al.}, ``Toward dual-functional radar-communication systems:
  Optimal waveform design,'' \emph{IEEE Transactions on Signal Processing},
  vol.~66, no.~16, pp. 4264--4279, 2018.

\bibitem{fascista2020low}
A.~Fascista \emph{et~al.}, ``Low-complexity accurate mmwave positioning for
  single-antenna users based on angle-of-departure and adaptive beamforming,''
  in \emph{IEEE International Conference on Acoustics, Speech and Signal
  Processing (ICASSP)}, 2020, pp. 4866--4870.

\bibitem{successiveLocBF_2019}
B.~Zhou \emph{et~al.}, ``Successive localization and beamforming in {5G} mmwave
  {MIMO} communication systems,'' \emph{IEEE Transactions on Signal
  Processing}, vol.~67, no.~6, pp. 1620--1635, 2019.

\bibitem{tasos_precoding2020}
A.~Kakkavas \emph{et~al.}, ``Power allocation and parameter estimation for
  multipath-based {5G} positioning,'' \emph{IEEE Transactions on Wireless
  Communications}, vol.~20, no.~11, pp. 7302--7316, 2021.

\bibitem{signalDesign_TVT_2022}
M.~F. Keskin \emph{et~al.}, ``Optimal spatial signal design for mmwave
  positioning under imperfect synchronization,'' \emph{IEEE Transactions on
  Vehicular Technology}, vol.~71, no.~5, pp. 5558--5563, 2022.

\bibitem{e2e_radar_TAES_2022}
W.~Jiang \emph{et~al.}, ``Joint design of radar waveform and detector via
  end-to-end learning with waveform constraints,'' \emph{IEEE Transactions on
  Aerospace and Electronic Systems}, vol.~58, no.~1, pp. 552--567, 2022.

\bibitem{learnProbingmmWave_2022}
Y.~Heng \emph{et~al.}, ``Learning site-specific probing beams for fast {mmWave}
  beam alignment,'' \emph{IEEE Transactions on Wireless Communications}, 2022.

\bibitem{Liu2022isac}
C.~Liu \emph{et~al.}, ``Learning-based predictive beamforming for integrated
  sensing and communication in vehicular networks,'' \emph{IEEE Journal on
  Selected Areas in Communications}, vol.~40, no.~8, pp. 2317--2334, 2022.

\bibitem{mateos2021end}
J.~M. Mateos-Ramos \emph{et~al.}, ``End-to-end learning for integrated sensing
  and communication,'' in \emph{IEEE International Conference on Communications
  (ICC)}, 2022, pp. 1942--1947.

\bibitem{Oshea2017intro}
T.~O’Shea \emph{et~al.}, ``An introduction to deep learning for the physical
  layer,'' \emph{IEEE Transactions on Cognitive Communications and Networking},
  vol.~3, no.~4, pp. 563--575, 2017.

\bibitem{Aoudia2018alternating}
F.~A. Aoudia \emph{et~al.}, ``End-to-end learning of communications systems
  without a channel model,'' in \emph{2018 52nd Asilomar Conference on Signals,
  Systems, and Computers}, 2018, pp. 298--303.

\bibitem{chen2022sdoanet}
P.~Chen \emph{et~al.}, ``{SDOAnet}: An efficient deep learning-based {DOA}
  estimation network for imperfect array,'' \emph{arXiv preprint
  arXiv:2203.10231}, 2022.

\bibitem{yassine2022fl}
T.~Yassine \emph{et~al.}, ``{mpNet}: Variable depth unfolded neural network for
  massive {MIMO} channel estimation,'' \emph{IEEE Transactions on Wireless
  Communications}, vol.~21, no.~7, pp. 5703--5714, 2022.

\bibitem{Zheng2017coupling}
Z.~Zheng \emph{et~al.}, ``Robust adaptive beamforming against mutual coupling
  based on mutual coupling coefficients estimation,'' \emph{IEEE Transactions
  on Vehicular Technology}, vol.~66, no.~10, pp. 9124--9133, 2017.

\bibitem{Fascista2019downlinkPos}
A.~Fascista \emph{et~al.}, ``Millimeter-wave downlink positioning with a
  single-antenna receiver,'' \emph{IEEE Transactions on Wireless
  Communications}, vol.~18, no.~9, pp. 4479--4489, 2019.

\bibitem{kingma2014adam}
D.~P. Kingma \emph{et~al.}, ``Adam: A method for stochastic optimization,''
  \emph{arXiv preprint arXiv:1412.6980}, 2014.

\bibitem{kay1993fundamentals}
S.~M. Kay, \emph{Fundamentals of statistical signal processing: estimation
  theory}.\hskip 1em plus 0.5em minus 0.4em\relax Prentice-Hall, Inc., 1993.

\bibitem{Chen-HWI-2022}
H.~Chen \emph{et~al.}, ``{MCRB}-based performance analysis of {6G} localization
  under hardware impairments,'' in \emph{IEEE International Conference on
  Communications Workshops (ICC Workshops)}, 2022, pp. 115--120.

\end{thebibliography}

\end{document}